\documentclass[mathleft,fleqn,%
]{an}
\usepackage{graphicx}
\usepackage[varg]{txfonts}
\def\ls{{_<\atop^{\sim}}}
\def\gs{{_>\atop^{\sim}}}
\begin{document}

\Pagespan{1}{}
\Yearpublication{2017}%
\Yearsubmission{2016}%
\Month{0}%
\Volume{999}%
\Issue{0}%
\DOI{asna.201400000}%

\title{A Decade of WHIM Searches: \\
        Where do we Stand and Where do we Go?}

\author{F. Nicastro\inst{1.2}\fnmsep\thanks{Corresponding author:
        \email{fabrizio.nicastro@oa-roma.inaf.it}}
\and  Y. Krongold\inst{3}
\and S. Mathur\inst{4}
\and M. Elvis\inst{2}
}
\titlerunning{Instructions for authors}
\authorrunning{T.\,H.\,E. Editor \& G.\,H. Ostwriter}
\institute{
Istituto Nazionale di Astrofisica - Osservatorio Astronomico di Roma, Monte Porzio Catone, 00078 RM, Italy 
\and 
Harvard-Smithsonian Center for Astrophysics, Cambridge, MA 02138, USA 
\and 
Instituto de Astronomia - Universidad Nacional Autonoma de Mexico, Mexico City, DF, Mexico
\and 
Department of Astronomy Ohio State University, Columbus, OH 43210, USA}

\received{15 Oct 2016}
\accepted{7 Nov 2016}
\publonline{XXXX}


\abstract{%
In this article we first review the past decade of efforts in detecting the missing baryons in the Warm Hot Intergalactic Medium (WHIM) 
and summarize the current state of the art by updating the baryon census and physical state of the detected baryons in the local Universe. 
We then describe observational strategies that should enable a significant step forward in the next decade, while waiting for the step-up in 
quality offered by future missions. 
In particular we design a multi-mega-second and multiple cycle XMM-{Newton} {\em legacy' program} (which we name the Ultimate Roaming Baryon Exploration, 
or URBE) aimed to secure detections of the peaks in the density distribution of the Universe missing baryons over their entire predicted range of temperatures.}

\maketitle

\section{Introduction}
At some point in the history of the Universe, baryons got lost. 
They were present and countable at the surface of last scattering (z=1000), when the Universe was only 380000 years old (Komatsu et al., 2009; Ade et al., 2015), and 
they were again there 2 billions years later (z=3), mostly in the Ly$\alpha$ Forest (Rauch 1998; Weinberg 1997). However, over the next 11 billion years, we have lost 
track of about half of them. During this complicated and restless puberty of the Universe, structures began forming copiously and baryons started shining in stars, 
in the intra-cluster medium and in active galactic nuclei (AGNs), becoming more and more directly visible and so countable. 
Yet today we can only account for $<60$\% of the predicted baryons; 
that is $>40$\% of the baryons are  now missing (e.g. Fukugita, 2003; Shull, Britton \& Danforth, 2012; Fig. 1). 
\begin{figure}
\hspace{-2cm}
\vspace{-1cm}
\includegraphics[width=100mm,height=100mm]{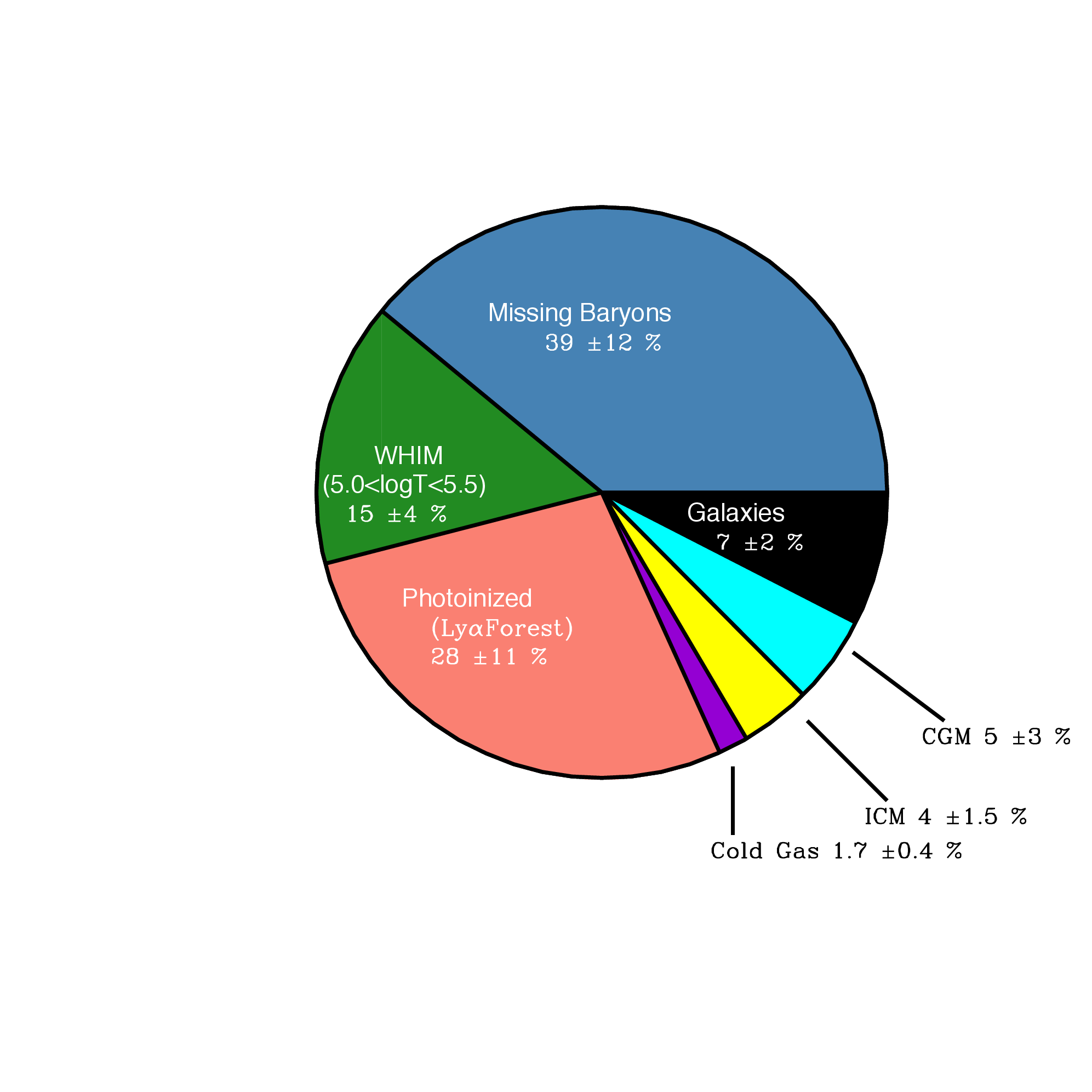}
\caption{Baryon budget in the Universe, at z=0. The actual percentage of baryons still missing (blue slice) could be as high as $\simeq 50$\%. The 5.0$<$logT$<$5.5 
WHIM (green slice), includes the WHIM independently detected (and double counted, see \S 3.1.1) in OVI and HI Ly$\alpha$ (Shull, Britton \& Danforth, 2012) and the 
cool X-ray WHIM (see \S 3.1.1).}
\label{fig1}
\end{figure}

What is perhaps even more embarrassing, today ($z\simeq 0$) baryons are missing on two very different scales, from the cosmological all the way down to the galactic: 
most galaxies fall short of 
baryons when their measurable baryonic fraction is compared with the universal baryon to total mass ratio (e.g. Ade et al., 2015) (Fig. 2). 
The problem is more severe for smaller galaxies, which suggests that lower potential well halos fail to retain baryons during mergers and/or bursts of star formation 
activity (e.g. Bell et al., 2003; McGaugh, 2010). 
If so, these baryons should be found in the galaxy Circum-Galactic Medium (CGM), at or even beyond their virial radii. 
\begin{figure}
\includegraphics[width=\linewidth,height=110mm]{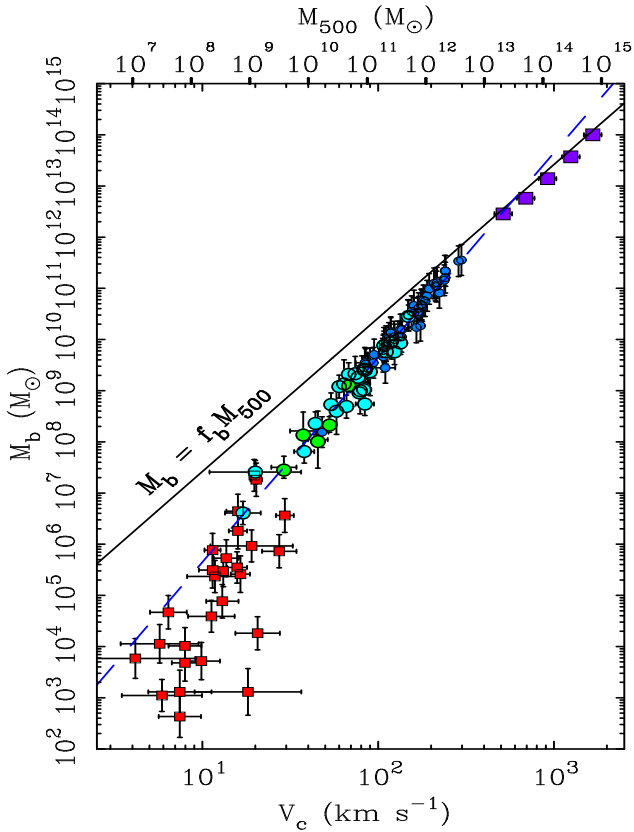}
\caption{Baryonic vs dynamical mass of virialized structures in the local Universe. From McGaugh et al. (2010).}
\label{fig2}
\end{figure}

These two missing-baryon problems are probably related and reconciled by the leading, $\Lambda$-CDM (Cold Dark Matter), Standard Cosmological Model (SCM), 
which predicts that these recycled baryons are hiding in a web of tenuous warm-hot intergalactic matter (the WHIM), whose nodes host the gaseous outskirts of formed 
structures, while continuously roaming in and out of them (e.g. Cen \& Ostriker, 2006). 

\section{Physics of the WHIM}
If this theory is right, then the baryons in the WHIM are hot (T$=10^5-10^7$ K) and tenuous (n$_b=10^{-6} - 10^{-4}$ cm$^{-3}$), and only couples with radiation 
through electronic transitions of residual HI (producing thermally broadened Ly$\alpha$ lines: BLAs) and light metals (C, N, O, Ne) in their Li- to H-like ionization 
stages. 
The strongest of these transitions fall in the soft X-ray band (e.g. Cen \& Fang, 2006) and should imprint narrow absorption lines in the X-ray spectra of distant 
astronomical sources. 
This is why, when 15 years ago the first two X-ray spectrometers with sufficient resolving power (albeit with small throughput) became available, the search for the WHIM 
finally started in earnest. 

\section{The WHIM: State of the Art}
\subsection{Circumgalactic Medium}
Copious evidence that our own Galaxy is permeated by and/or embedded in a bath of warm-hot material, was soon found using the {\em Chandra} and XMM-{\em Newton} 
grating spectrometers (Nicastro et al., 2002; 2003). 
The existence of this matter has been confirmed many times since and it is not questioned (Williams et al., 2005; Wang et al., 2005; Anderson \& Bregman, 2010; 
Gupta et al., 2012; Fang et al., 2013, 2015). Only recently though has this matter been finally located. It lies in a large spherical volume centered on the Galaxy's center, 
with a radius $\ge 70-200$ kpc. Its mass is estimated to be sufficient to finally close the Galaxy's baryon census (Nicastro et al., 2016a; Gupta et al., 2012). 

\subsection{Intergalactic Medium}
While these results locate and characterize the majority of the baryons in and around our own Galaxy, on the large, cosmological, scale the hunt for the WHIM remains 
difficult. 
This is because the absorption signal expected from the WHIM is extremely weak: only 1-2 strong OVII He$\alpha$ (i.e. $1s^2 \rightarrow 1s2p$) absorbers, 
with EW$\ge 20$ m\AA\ (requiring 
$\ge 120$ counts per $\Delta\lambda=70$ m\AA\ resolution element, to be detected with, e.g., the XMM-{\em Newton} RGS), are predicted to exist per unit redshift. 
More typical (dN/dz$\sim 10$) OVII He$\alpha$ absorbers have EW$\sim 3$ m\AA\ (e.g. Cen \& Fang, 2006). 
Such EWs are very difficult to be detected uncontroversially. This is not only because they require 5000 counts per 70 m\AA\ for a 3$\sigma$ detection, but also because 
the contrast EW/$\Delta\lambda = 4$\% approaches perillously our knowledge of the systematics of the current 
spectrometers. These are $\simeq 3$\% for the {\em Chandra} LETG (J. Drake, 2016, private communication), and $\sim2$\% for the XMM-{\em Newton} RGSs (Kaastra, 2016), 
in the relevant band pass. 

These limitations call for demanding observing strategies: very bright background targets at the highest possible redshift, very long integrations, and/or the 
selection of direct or indirect WHIM signposts, or all of the above. In addition at least two independent confirmations of a proposed detection are needed, when the 
spectrometer systematics are close to the detection threshold (i.e. for line EW$\simeq 1-3$ m\AA). 

For example, more than 10 years ago, we proposed to observe blazars in powerful outburst states to obviate the spectrometers’ small throughput. 
In practice, however, this turned out to be feasible only for the luminous and nearby blazar Mkn~421, whose 2003 {\em Chandra} observations yielded the highest S/N 
high resolution X-ray spectrum ever taken. 
The analysis of this spectrum led in 2005 to the first two claims of WHIM detections (Nicastro et al., 2005a,b). As science imposes, these claims were strictly scrutinized 
and soon became highly controversial (Rassmussen et al., 2007; Kaastra et al., 2006; Nicastro, Mathur \& Elvis, 2008). 
Since then, there have been new tentative discoveries reported. However, they are either serendipitous (Nicastro et al., 2010; Zappacosta et al., 2012) 
or trace extreme galaxy over-densities (Buote et al., 2009; Fang et al., 2010, Ren, Fang \& Buote, 2015). In both cases sample only the hottest and densest gas 
and are not representative of the majority of the WHIM. 
Moreover, in two of these cases (Buote et al., 2009; Fang et al., 2010, Ren, Fang \& Buote, 2015), 
the WHIM identification has been challenged by an alternative identification of the absorption line with a cold Galactic medium (Nicastro et al., 2016b). 

\subsection{The Best WHIM Target in the Universe}
In 2010 we identified what we define ``the best WHIM target in the Universe'': the relatively high redshift ($z>0.4$; Danforth et al., 2010) and X-ray bright 
(F$_{0.3-2 keV} \simeq (1-20)\times 10^{-11}$ erg cm$^{-2}$ s${-1}$: Fig. 1) blazar 1ES~1553+113. 

We coordinated a strong observational effort to study this source, which started in 2010, when a high S/N HST-COS spectrum of the blazar 1ES 1553+113 was taken. 
This observation set a lower limit of $z>0.4$ for the redshift of 1ES~1553+113 (making it the X-ray brightest quiescent extragalactic source in the $z\ge 0.4$ sky: 
Fig. 3). The COS spectrum showed 5 possible BLAs (possible tracers of the cool end - T$\simeq 10^{5.2}-10^{5.5}$ K - of the WHIM mass distribution). 

A follow up 0.5 Ms {\em Chandra} LETG observation of the target, in 2012, detected 
3 CV-BLA and 1 OVII-BLA associations (Nicastro et al., 2013), establishing a fraction of cool WHIM of 15\%(Fig. 1) and so halving the Lyα+OVI fraction in Shull, Britton \& 
Danforth (2012). 
Finally, in the summer of 2015 the first half of a 1.6 Ms total XMM-{\em Newton} observation of 1ES~1553+113 was performed (\S 3.1.2). 
\begin{figure}
\includegraphics[width=\linewidth,height=90mm]{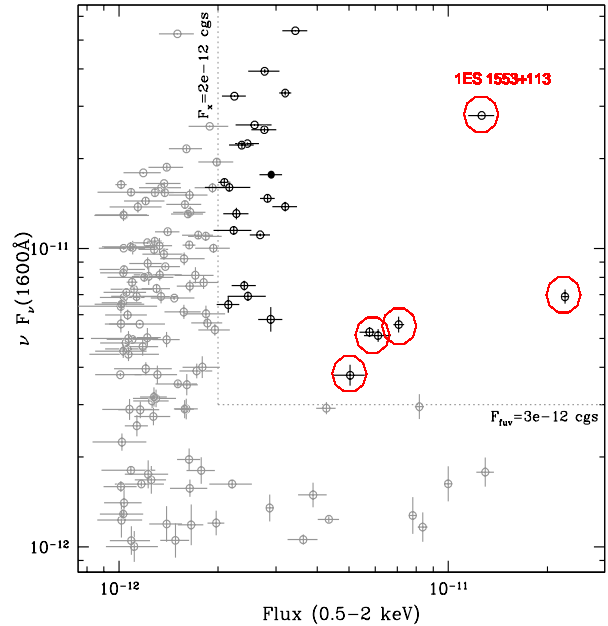}
\caption{All known AGNs at $z>0.3$ and with FUV (1600 \AA) and X-ray (0.5-2 keV) fluxes larger than $10^{-12}$ erg s$^{-1}$ cm$^{-2}$ (i.e. 0.05 mCrab). 
The insert defines the region of the 26 $z>0.3$ targets with $\nu F_{\nu}(1600)\ge 3\times 10^{-12}$ erg s$^{-1}$ cm$^{-2}$ and $F(0.5-2 keV) \ge 0.1$ mCrab, 
accessible by both the HST-COS and, in the future, the {\em Athena}-XIFU. The 6 red circles mark the 6 objects accessible now with the HST-COS and the 
XMM-{\em Newton} and {\em Chandra} gratings.}
\label{fig3}
\end{figure}

\subsubsection{Physical State and Mass of the $5.0 <$logT$<5.5$ WHIM}
The 2012 500 ks {\em Chandra} observation of 1ES 1553+113 yielded a spectrum that is 3$\sigma$-sensitive to CV column densities about 2-3 times smaller than 
OVII columns (and both still far from the known level of systematics of the spectrometer: Fig. 4), in the $\le 0.4$ redshift range. 
This is because the LETG throughput stays roughly constant from 21.6 \AA\ (the rest frame wavelength of the 
OVII He$\alpha$ transition) up to 56.7 \AA\ (the position of the CV He$\alpha$ transition at $z=0.4$), while the resolving power of the spectrometer increases linearly 
over the same spectral range. The result is to rough halving the line equivalent width detectability threshold. 

Moreover the BLA signposts discovered by COS along the line of sight to 1ES~1553+113 (Danforth et al., 2010; Nicastro et al., 2013) have Doppler parameters 
$b\ls 80$ km s$^{-1}$, so can only trace gas with $5.0 \ls$logT$\ls 5.5$, where the fraction of CV is typically higher than the fraction of OVII. 
The LETG spectrum detected 3 CV-BLA possible associations and a putative OVII-BLA association (Nicastro et al., 2013), whose temperature, 
equivalent H column density, baryon volume density and metallicity are listed in Table 1. Temperature and volume densities are derived by fitting the X-ray spectra 
with our hybrid-ionization (collisional ionization plus the meta-galactic photo-ionization contribution: Nicastro et al., 2005b) models. For all the 4 systems the 
estimated temperatures (constrained mainly by the absence of detectable OVII absorption associated with the CV filaments and the absence of detecable OVIII/NeIX 
for the putative OVII filament) are in agreement with the thermal broadening measured for the associated BLAs. Volume densities are only roughly estimated 
via the relative importance of photoionization versus collisional ionization: at n$_b \gs 100$ cm$^{-3}$ the gas is virtually collisionally ionized and while the fraction of 
CV is close to unity, little or no OVII is present at logT$\ls 5.3$; conversely, at n$_b \ls 10$ cm$^{-3}$ and 5.0$\ls$logT$\ls$5.5, the photoionization contribution 
depopulate CV while boosting the fraction of OVII without significanly modifying (compared to collisional ionization equilibrium) the fractions of OVIII and NeIX. 
\begin{figure}
\includegraphics[width=\linewidth,height=90mm]{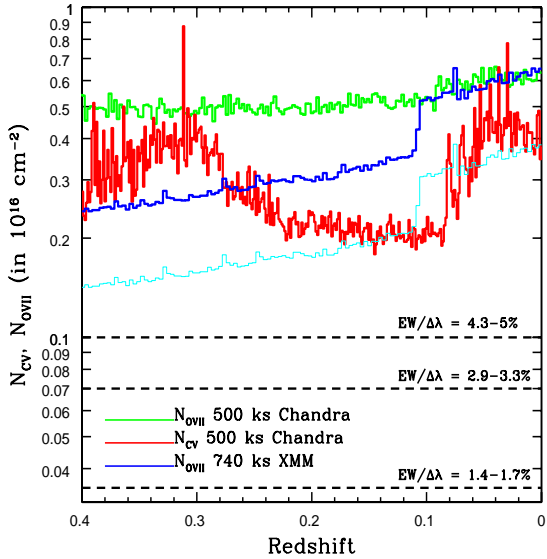}
\caption{CV and OVII column density 3$\sigma$ detectability threshold as function of redshift, for the existing 500 ks {\em Chandra}-LETG 
(green and red curves) and 740 ks XMM-{\em Newton}-RGS (blue curve: CV is external to the RGS band) spectra of 1ES~1553+113. 
The light cyan curve marks the predicted detectability of the total XMM-{\em Newton} RGS spectrum, once the full 1.6 Ms observation will have been completed. 
The three horizontal dashed lines, mark the column densities corresponding to line EW$\simeq 1$, 2 and 3 m\AA, down to which EW to spectral resolution contrasts 
are as labeled, with the lower and upper boundaries corresponding to the XMM-{\em Newton} RGS and the {\em Chandra} LETG), respectively. 
These should be compared with our current $\sim 3$\% knowledge of the systematics of the current spectrometers (see test). 
}
\label{fig4}
\end{figure}
\begin{table*}
\caption{Physics and Chemistry of the 3 CV-BLA and 1 OVII-BLA Associations along the line of sight to 1ES~1553+113.}
\label{tab1}
\begin{tabular}{ccccc}\hline
Redshift & logT & N$_b$ & n$_b$ & Z/Z$_{\odot}$ \\ 
& (T in k) & (in $10^{19}$ cm$^{-2}$) & (in $10^{-6}$ cm$^{-3}$) & \\
\hline
$0.041 \pm 0.002$ & $5.45 \pm 0.05$ & $3.8 \pm 0.8$ & 1 & $0.13^{+0.07}_{-0.10}$ \\
$0.190 \pm 0.002$ & $5.25 \pm 0.05$ & $1.9 \pm 1.8$ & 107 & $0.4^{+0.4}_{-0.3}$ \\
$0.237 \pm 0.002$ & $5.03 \pm 0.01$ & $0.3^{+0.2}_{-0.1}$ & 109 & $2.7^{+1.8}_{-2.2}$ \\
$0.312 \pm 0.002$ & $5.25 \pm 0.05$ & $3.1^{+1.6}_{-1.1}$ & 112 & $0.32^{+-0.19}_{-0.22}$ \\
\hline
\end{tabular}
\end{table*}
The weighted average of the metallicity of the 4 systems is Z$=0.3$  Z$_{\odot}$, in agreement with theoretical predictions (e.g. Cen \& Fang, 2006). 

Given the estimated metallicity and ionization (temperature) corrections, we can use these four systems to estimate the fraction of cosmological mass density 
of baryons contained in IGM gas with temperatures $5.0 <$logT$<5.5$, independent of the particular ion species considered. We get $\Omega_b(5.0<$logT$<5.5) = 
0.0069 \pm 0.0018$. I.e. a fraction $(15 \pm 4)$\% of $\Omega_b$ derived from fluctuations of the cosmic microwave background (e.g. Komatsu et al., 
2009; Ade et al., 2015). 

This baryon fraction is virtually identical to the ones estimated by Shull et al. (2012) independently for the two ions HI (BLAs: $\Omega_b(BLA) 
= 14 \pm 7)$ and OVI ($\Omega_b(OVI) = 17 \pm 4$), both tracing the cool end of the WHIM. We therefore conclude that BLAs (HI) and OVI trace the same gas as the 
BLA-CV/OVII associations. Hence counting them separately, without properly account for metallicity and ionization corrections, leads to double-counting the 
cool-WHIM baryon contribution to the total baryon census. BLAs, OVI and CV, all trace the same WHIM gas in the $5.0<$logT$<5.5$ temperature range, and contribute 
only $(15 \pm 4)$\% to the baryon census in the local Universe (Fig. 1). 

This result implies that up to $\sim 50$\% of the baryons are still missing (Fig. 1). 
They are expected to lie in hotter IGM gas, at temperature logT$>5.5$, best traced by the He-like and H-like ions of oxygen (OVII and OVIII).  

\subsubsection{The XMM-{\em Newton} Very Large Program} 
For this reason in 2015 we proposed that the XMM-{\em Newton} RGS, with an effective area 2.5-5 times larger than the {\em Chandra} LETG in the OVII He$\alpha$ 
$z<0.4$ redshift range, observed 1ES~1553+113 for 1.6 Ms. The program was approved, and the first 760 ks of the observation were taken during the first 
XMM-{\em Newton} cycle 14 visibility window (beetween July and September 2015). Unfortunately,  during this first half of the full observing campaign, the target was 
caught at a flux level of only 0.3 mCrab ($0.6 \times 10^{-11}$ erg s$^{-1}$ cm$^{-2}$) in the 0.3-2 keV band, close to its historical minimum. 
1ES~1553+113 can be even a factor of 20 brighter than this and a large as possible total number of photons is vital to detect weak 
OVII WHIM absorption lines. So, we postponed the second part of the XMM-{\em Newton} observation of 1ES~1553+113 to the first trimester of 2017, when, 
according to a periodicity claim for this source (The Fermi Collaboration, 2015), the target should reach again levels of -10 mCrab. 

The first part of the XMM-{\em Newton} RGS observation of 1ES~1553+113 did not yield conclusive results. The RGS spectrum is currently 3$\sigma$-sensitive to 
OVII column densities of $\sim 2.5\times 10^{15}$ cm$^{-2}$ at $z>0.1$ and about twice that at $z<0.1$ (Fig. 4, blue curve). 
These correspond to unsaturated line equivalent width sensitivities  EW$\gs 7-14$ m\AA, which are factors of $\simeq 2-4$ above the conservative 
3\% spectrometer systematics (Fig. 4). 
According to the most conservative predictions to date (Cen \& Ostriker, 2006), about 2-5 OVII He$\alpha$ absorbers per unit redshift are expected along a random 
WHIM line of sight, down to these EWs. Given the $z\ge 0.4$ redshift of 1ES~1553+113, we would expect to detect about 1-2 OVII He$\alpha$ intervening 
absorption line at single-line statistical significance $\ge 3\sigma$ in the current RGS spectrum of 1ES~1553+113.

The current RGS spectrum confirms, at the same (low) single-line statistical significance (2.2$\sigma$), the $z=0.043$ OVII He$\alpha$ line hinted at 
in the LETG spectrum. In addition and a new, hotter (logT$=5.7^{+0.5}_{-0.1}$), OVII system is possibly detected in two transitions (He$\alpha$ and He$\beta$ - i.e. 
$1s^2 \rightarrow 1s3p$). A consistency check with the LETG spectrum does not rule out the detection (Fig. 5). 
\begin{figure}
\includegraphics[width=\linewidth,height=60mm]{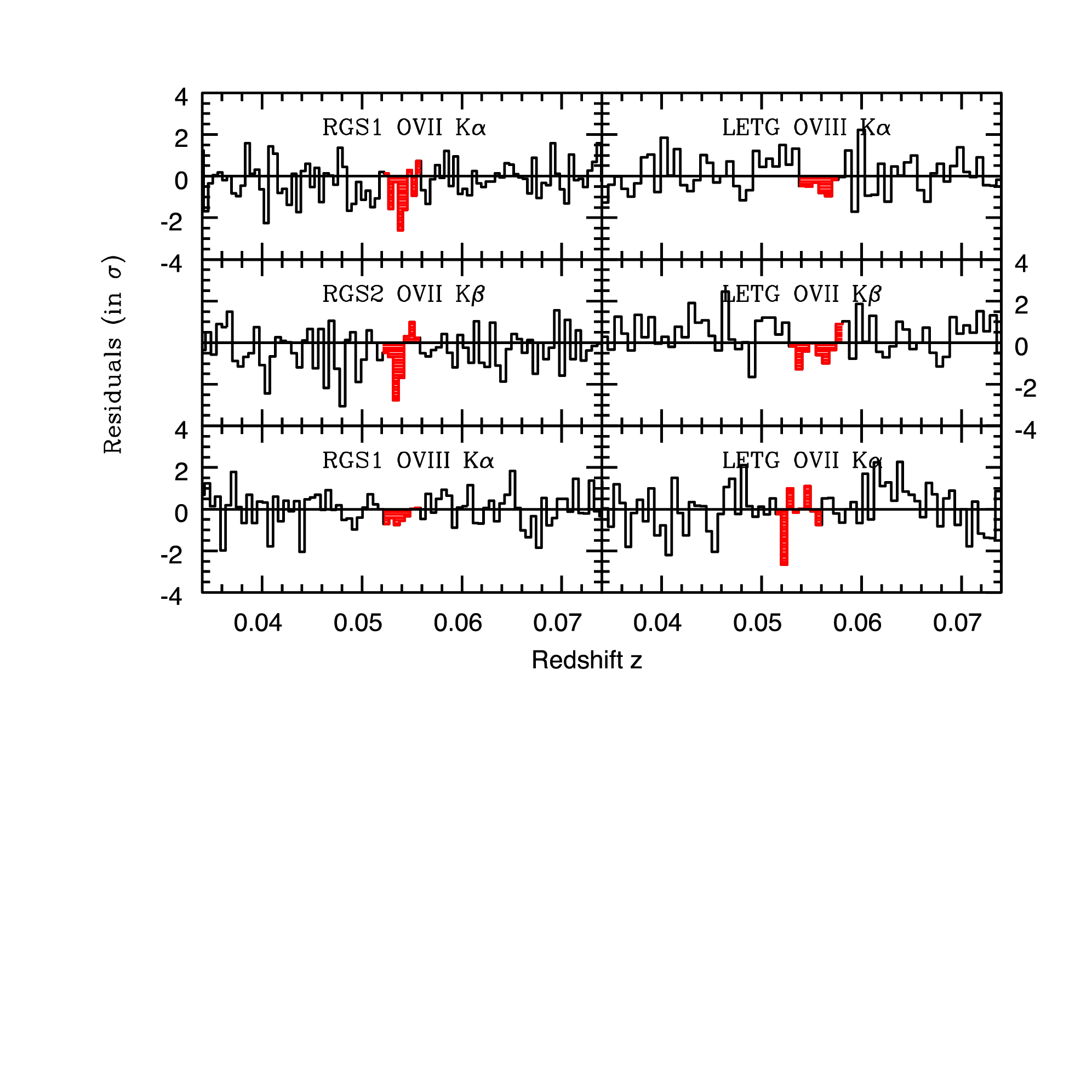}
\caption{Portions of the existing 740 ks XMM-{\em Newton} RGS (left panels) and 500 ks {\em Chandra} LETG (right panels) spectra of 1ES~1553+113 
around the $z=0.054$ positions of the OVII He$\alpha$, OVII K$\beta$ and OVIII Ly$\alpha$ transitions.}
\label{fig5}
\end{figure}
The current RGS spectrum of 1ES~1553+113 is thus still consistent with theoretical predictions (especially after factoring the large uncertanties in the predicted 
number density of absorbers due to cosmic variance), but significant limits are close to being reached. A continued lack of detections will require a revised theoretical 
understanding  of the WHIM phenomenon.

\section{Conclusions and Future of WHIM Studies}

\subsection{Next Decade with XMM-{\em Newton} and {\em Chandra}}  
A firm, high-significance detection of the majority of the missing baryon mass, at temperatures logT$>5.5$ has not yet been secured and so, despite sytong efforts 
and controversies, the missing baryons continue to hide. 
Hopefully the completion of the XMM-{\em Newton} Very Large Program on the best WHIM target in the Universe, by halving the 3$\sigma$ OVII column density 
sensitivity (cyan curve in Fig. 4), will finally secure this first uncontroversial detection (or start seriously questioning WHIM predictions). 
However, even the dear detection of one or two systems along a single line of sight is obviously not sufficient to provide a reasonable estimate of the cosmological 
mass density of the WHIM in the temperature interval $5.5 <$logT$<6.7$, and self-evidently cannot address cosmic variance. Tthe number density of WHIM filaments 
with column density larger than a given threshold, is expected to vary significantly from line of sight to line of sight, e.g. Cen \& Fang, 2006. 

More lines of sight must be investigated to the same depth to which is being investigated the line of sight to the blazar 1ES~1553+113. I.e. a factor of $\sim 2$ above 
the spectrometer systematic: cyan curve in Fig. 4. 
This should be done by both FUV (i.e. the HST-COS) and X-ray (the XMM-{\em Newton} and {\em Chandra} spectrometers), to guarantee the detection of 
both metals (without which a proper ionization correction is impossible) and HI (without which a proper metallicity correction is impossible). 

This program requires that the background targets be not only sufficiently distant ($z>0.3$), but also bright enough both in the FUV and X-rays, to guarantee a 
Signal to Noise per Resolution Element (SNRE) in the continuum sufficient to detect weak OVII (EW=5-7 m\AA) and BLA (EW=10-30 m\AA) absorption lines, but 
still far (factors of 2-3) from our knowledge of the instrument systematics. 
Fig. 3 shows a plot of all known type-1 AGNs at $z>0.3$ (all points, grey and black) with FUV (1600 \AA) and X-ray (0.5-2 keV) fluxes larger than $10^{-12}$ erg 
s$^{-1}$ cm$^{-2}$. The dotted lines define the region of targets with both $\nu F_{\nu}(1600)\ge 3\times 10^{-12}$ erg s$^{-1}$ cm$^{-2}$ and 
$F(0.5-2 keV) \ge 2\times 10^{-12}$ erg s$^{-1}$ cm$^{-2}$, accesible by both the HST-COS and, in the future, the {\em Athena}-XIFU. 

In this flux-flux diagram we also identify 6 targets (red circles in Fig. 3) that are sufficiently bright to be observed within the next decade and with a moderately strong 
effort in terms of total exposure time. All these 6 targets have already HST-COS spectra, but a total of 150 additional HST orbits would be required to reach the needed 
sensitivity to $70 \ls b\ls 140$ km s$^{-1}$ (i.e. 5.5$\ls$logT$\ls 6.0$) BLAs. 

We sugest that the X-ray part of this program should be defined as an XMM-{\em Newton} {\em Legacy}-WHIM program. 
One of these 6 targets is 1ES~1553+113 (Fig. 3), for which both {\em Chandra} and XMM-{\em Newton} spectra are already sufficiently deep for WHIM studies. 
The additional five targets cover a total potential WHIM path-length of $\Delta z = 2.24$ and need a total of 11 Ms of XMM-{\em Newton} 
time to be covered at the same depth level as the $z<0.4$ pathlength towards 1ES~1553+113. 
Excluding the FUV/X-ray faintest target, would reduce the required X-ray exposure to 8.1 Ms. 
It would be prudent if these five observations be fully carried out only after a first shallower (250-500 ks per target, depending on the targets) campaign of the 
5 targets has been completed. Such pilot spectra will be $3\sigma$ sensitive to N$_{OVII} \ge 5 \times 10^{15}$ cm$^{-2}$, down to which about 2-3 OVII absorbers 
are predicted per unit redshift (Cen \& Fang, 2006). This, factoring both the total $\Delta z = 2.24$ pathlength and uncertainties due to small-number statistics 
(as an attempt to account for cosmic variance; Gehrels, 1986), should guarantee both the detections of at least 2 strong systems (probably associated with the WHIM web virialized nodes) and a 
good first assessment of cosmic variance. Additionally, exploring first these 5 lines of sight at a shallower level than fully needed to guarantee the detection of at least 
one system per line of sight and so an $\Omega_b(5.5<$logT$<6.7)$ estimate together with a thorough assessment of cosmic variance, will also possibly allow us to 
reduce both the number and depth of subsequent follow ups, which will clearly depend on the output of the pilot campaign.

\noindent
UV (provided by the Optical-Monitor of XMM-{\em Newton}) as well as ground-based Optical, IR and mm follow-ups of the WHIM signposts detected along these 
6 lines of sight, will be essential to study the galaxy environment of WHIM filaments and better understand the interplay bvetween virialized and unviriliazed structures 
in the Universe. 

\subsection{The Future with the {\em Athena}-XIFU}
Starting from 2028, the X-Ray Integral Field Unit Spectrometer (XIFU: Barret et al., 2016) of {\em Athena}, will open the way to deep studies of the WHIM. 
{\em Athena} will routinely study hundreds of WHIM filaments along several tens of different lines of sight at $z<2$ (Fig. 6 and 7), and will enable: 
(1) accurate (few \%) measurement of the cosmological mass density of baryons in the Universe; (2) physical, kinematical and chemical studies of the low-z IGM. 
In synergy with future FUV, O/IR and mm instrumentation will allow us to: (3) detect rare metal isotopes in {\em Athena} WHIM priors and possibly molecules 
in their cold interfaces (e.g. Sunyaev \& Churazov, 1987;  Sunyaev \& Docenko, 2007); (4) identify WHIM-galaxy associations and map the structure of galaxy 
clustering; (5) study the interplay between galaxy and AGN outflows and the IGM (feedback); (6) understand the role of shocks in the formation of 
structures in the Universe. 
\begin{figure}
\includegraphics[width=\linewidth,height=80mm]{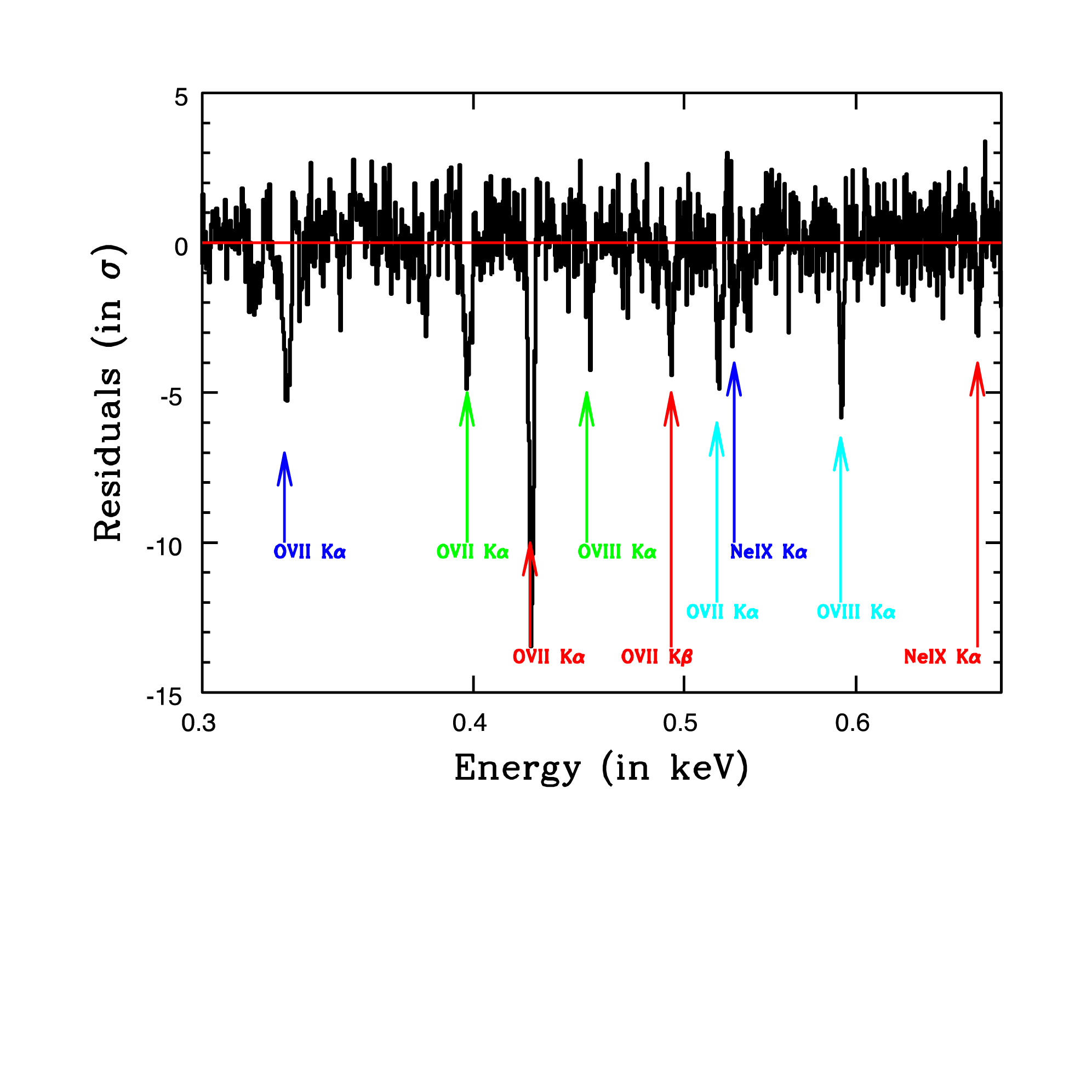}
\caption{0.3-0.7 keV portion of a mock {\em Athena}-XIFU spectrum of the brightest known blazar at $z\ge 0.8$ along a random WHIM line of sight as predicted by 
the Cen \& Ostriker (2006) hydrodynamical simulations. Four WHIM filaments are clearly detected at single-line statistical significances $>5\sigma$, all in multiple lines.}
\label{fig6}
\end{figure}
\begin{figure}
\includegraphics[width=\linewidth,height=80mm]{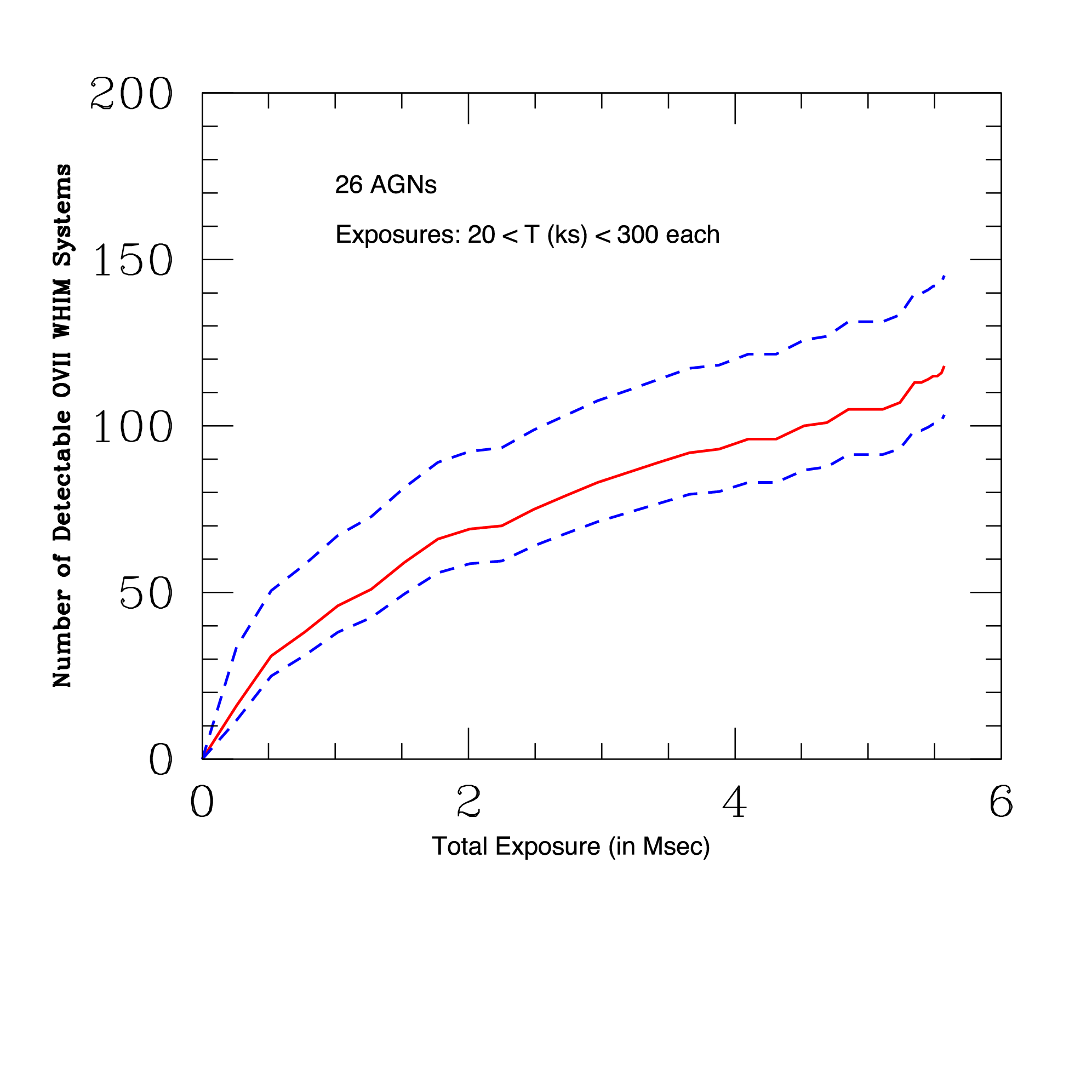}
\caption{Predicted number of OVII WHIM filaments detectable with the {\em Athena}-XIFU against the 26 known brightest AGNs at $z>0.3$ (Fig. 3, black points), as a function 
of total observing time (solid red curve), and its 90\% uncertainties (dashed blue curves).}
\label{fig7}
\end{figure}
%

\subsection{References}
Ade, P.A.R. et al., 2015, arXiv:1502.01589 \\
Anderson, M.E. \& Bregman, J.N., 2010, ApJ, 714, 320. \\ 
Barret, D. et al., 2016, SPIE, 9905, in press (arXiv:1608.08105) \\
Bell, E.F. et al., 2003, ApJ, 585, L117. \\
Buote, D.A. et al., 2009, ApJ, 695, 1351. \\
Cen, R. \& Ostriker, J.P., 2006, ApJ, 650, 560. \\
Cen, R. \& Fang, T., 2006, ApJ, 650, 573. \\
Danforth, C.W. et al., 2010, ApJ, 710, 613. \\
Fang, T. et al., 2015, ApJS, 217, 21. \\ 
Fang, T. et al., 2013, ApJ, 762, 20. \\
Fang, T. et al., 2010, ApJ, 714, 1715. \\
The Fermi Collaboration, 2015, ApJ, 813, L41. \\
Fukugita, M., 2003, IAU sym. 220, astro-ph/0312517. \\
Gehrels, N., 1986, ApJ, 303, 336. \\
Gupta, A. et al., 2012, ApJ, 756, L12. \\
Kaastra, J., 2016, these proceedings, in press. \\
Kaastra, J. et al., 2006, ApJ, 652, 189. \\
Komatsu, E., et al. 2009, ApJS, 180, 330. \\
McGaugh et al., 2010, ApJ, 708, L14. \\
Nicastro, F. et al., 2016a, ApJ, 828, L12 \\
Nicastro, F. et al., 2016b, MNRAS, 485, L123. \\
Nicastro, F. et al., 2013, ApJ, 769, 90. \\
Nicastro, F. et al., 2010, ApJ, 715, 854. \\
Nicastro, F. et al., 2008, Science, 319, 55. \\ 
Nicastro, F. et al., 2005a, Nature, 433, 495. \\
Nicastro, F. et al., 2005b, ApJ, 629, 700. \\
Nicastro, F., et al., 2003, Nature, 421, 719. \\ 
Nicastro, F., et al., 2002, ApJ, 573, 157. \\
Rasmussen, J. et al., 2007, ApJ, 656, 129. \\ 
Rauch, M., 1998, ARA\&A, 36, 267. \\
Ren, B., Fang, T. \& Buote, D.A., 2014, ApJ, 782, L6. \\ 
Shull, J.M. et al., 2012, ApJ, 759, 23. \\
Sunyaev, R.A. \& Docenko, D.O., 2007, AstL, 36, 67. \\
Sunyaev, R.A. \& Churazov, E., 1984, SvAL, 10, 201. \\
Wang, Q.D. et al., 2005, ApJ, 635, 286. \\
Weinberg, D.H., 1997, ApJ, 490, 564. \\
Williams, R. et al., 2005, ApJ, 631, 856. \\
Zappacosta, L. et al., 2012, ApJ, 753, 137. \\
Zappacosta, L. et al., 2010, ApJ, 717, 74. \\

\end{document}